\documentclass[pra,twocolumn, superscriptaddress, 10pt]{revtex4-2}
\usepackage{amssymb, amsmath,color,mciteplus,graphicx,subfigure}
\usepackage{calc,epsfig,epstopdf,mciteplus,bm,mathrsfs}
\usepackage{times}
\usepackage{float}
\usepackage{multirow}
\usepackage{lipsum}
\usepackage[ruled,vlined,linesnumbered]{algorithm2e}
\usepackage{dcolumn}
\usepackage{setspace}
\usepackage{xcolor}

\begin{document}
	
	\title{ Variational quantum algorithms for scanning the complex spectrum of non-Hermitian systems}
	
	\author{Xu-Dan Xie}
	\affiliation{Key Laboratory of Atomic and Subatomic Structure and Quantum Control (Ministry of Education),  Guangdong Basic Research Center of Excellence for Structure and Fundamental Interactions of Matter, and School of Physics, South China Normal University, Guangzhou 510006, China}
	
	\author{Zheng-Yuan Xue}  \email{zyxue@scnu.edu.cn}
	\affiliation{Key Laboratory of Atomic and Subatomic Structure and Quantum Control (Ministry of Education),  Guangdong Basic Research Center of Excellence for Structure and Fundamental Interactions of Matter, and  School of Physics, South China Normal University, Guangzhou 510006, China}
	\affiliation{Guangdong Provincial Key Laboratory of Quantum Engineering and Quantum Materials,  Guangdong-Hong Kong Joint Laboratory of Quantum Matter, and Frontier Research Institute for Physics,\\ South China Normal University, Guangzhou 510006, China}
	
	\author{Dan-Bo Zhang} \email{dbzhang@m.scnu.edu.cn}
	\affiliation{Key Laboratory of Atomic and Subatomic Structure and Quantum Control (Ministry of Education),  Guangdong Basic Research Center of Excellence for Structure and Fundamental Interactions of Matter, and  School of Physics, South China Normal University, Guangzhou 510006, China}
	\affiliation{Guangdong Provincial Key Laboratory of Quantum Engineering and Quantum Materials,  Guangdong-Hong Kong Joint Laboratory of Quantum Matter, and Frontier Research Institute for Physics,\\  South China Normal University, Guangzhou 510006, China}
	
	\date{\today}

	\begin{abstract}
		Solving non-Hermitian quantum many-body systems on a quantum computer by minimizing the variational energy is challenging  as the energy can be complex. Here, we propose a variational quantum algorithm for solving the non-Hermitian Hamiltonian by minimizing a type of energy variance, where zero variance can naturally determine the eigenvalues and the associated left and right eigenstates. Moreover, the energy is set as a parameter in the cost function and can be tuned to scan the whole spectrum efficiently by using a two-step optimization scheme. Through numerical simulations, we demonstrate the algorithm for preparing the left and right eigenstates, verifying the biorthogonal relations, as well as evaluating the observables. We also investigate the impact of quantum noise on our algorithm and show that its performance can be largely improved using error mitigation techniques. Therefore, our work suggests an avenue for solving non-Hermitian quantum many-body systems with variational quantum algorithms on near-term noisy quantum computers.
	\end{abstract}
	
	\maketitle
	\definecolor{RED}{RGB}{255,0,0}
	\section{Introduction} \label{sec:level1}
	The exploration of non-Hermitian physics holds great importance in physics, as numerous natural physical phenomena exhibit non-Hermitian characteristics~\cite{ashida2020non}. Recently, much attention  has been paid    to non-Hermitian physical systems due to their exceptional properties, including PT symmetry breaking~\cite{PhysRevLett.103.093902,bender2007making,dorey2001spectral,ozdemir2019parity}, skin effect~\cite{PhysRevLett.121.086803,PhysRevB.105.054315,PhysRevB.100.054301}, topological properties~\cite{PhysRevLett.108.220401,zhang2020non,PhysRevA.103.033325,PhysRevA.101.063612} and so on~\cite{heiss2004exceptional,PhysRevLett.123.170401,PhysRevE.59.6433,PhysRevLett.110.050403}. Nevertheless, tackling the many-body non-Hermitian physical systems presents a challenge  for classical computers due to the exponential growth of Hilbert space~\cite{barahona1982computational}. While tensor network provides a potential powerful method to solve non-Hermitian systems~\cite{PhysRevLett.130.100401,jaschke2018one,PhysRevB.98.235148}, it is imperative to limit the matrix dimension to a moderate threshold value to alleviate computational complexity\cite{orus2014practical}.
	
	Quantum computing has the potential to provide solutions for  hard problems~\cite{wiesner1996simulations,PhysRevLett.102.130503,PhysRevLett.79.2586,smith2019simulating}.  For near-term quantum computers, variational quantum eigensolver~(VQE) is designed to solve eigenstates of many-body systems~\cite{peruzzo2014variational,kandala2017hardware}. It can efficiently calculate the ground state~\cite{tilly2022variational, fedorov2022vqe,PhysRevX.6.031007,bosse2022probing,PhysRevB.106.214429} and low-lying excited states~\cite{PhysRevResearch.1.033062,higgott2019variational,PhysRevB.107.024204,xie2022orthogonal} of a given Hamiltonian. The efficacy of VQE algorithms is based on minimizing the variational energy~\cite{tilly2022variational}. However,  non-Hermitian systems may not have a minimum energy as the eigenvalue of the Hamiltonian may be a complex number, rendering the conventional VQE not implementable. Alternatively, the energy variance can be utilized as the cost function, as it can be used to determine an eigenstate since any eigenstate is characterized by zero energy variance~\cite{zhang2022variational,Chen_2023}. This approach allows us to bypass the energy minimization issue in non-Hermitian Hamiltonians~\cite{guo2022variational}. Therefore, energy variance can be leveraged as a cost function to design the variational quantum algorithm to effectively solve the non-Hermitian Hamiltonian.
	
	Here, we develop a variational quantum algorithm for solving non-Hermitian quantum systems by using a cousin of the energy variance as the cost function. The cost functon differs from the conventional energy variance in that the energy is parameterized with a complex number. By zero-variance principle, the eigenstates and the corresponding  eigenenergies can be obtained and self-verified by minimizing the energy variance to zero. For efficient optimization, we adopt a two-step optimization strategy, which allows the system to evolve towards the target quantum state by adjusting the optimization sequence among parameters. The effectiveness of the algorithm 
	is demonstrated through numerical simulations. Additionally, we investigate the impact of quantum noise on the algorithm and improve its performance by incorporating error mitigation techniques.
	Therefore, our work highlights the significant potential of quantum variational algorithms  for simulating non-Hermitian physics.
	
	The rest of this paper is organized as follows. In Sec.~\ref{sec:VQE_nonher}, we first introduce the non-Hermitian Hamiltonian in brief, and then propose the variational quantum algorithm as well as the optimization. In Sec.~\ref{sec:sumulaton}, we present the results of the numerical simulation of the algorithm, analyze the effects of quantum noise, and show the improvements using error mitigation. Finally, we draw conclusions in Sec.~\ref{sec:conclusion}.

	\section{\label{sec:VQE_nonher}Exploring THE QUANTUM VARIATIONAL ALGORITHM in Non-Hermitian Systems} 
	In this section, we initially provide a concise introduction to the non-Hermitian systems and the difficulties in utilizing the conventional VQE algorithm in this scenario. Then, we present a quantum variational algorithm that is suitable for the non-Hermitian Hamiltonian. Afterwards, a two-step optimization strategy is proposed to determine the desired eigenstates and eigenvalues. Finally, we show how to estimate operator expected values for non-Hermitian systems.
	
	\subsection{Motivation}
	In an isolated quantum system, Hermiticity is a fundamental postulate in the framework of quantum mechanics. This property guarantees that the expectation value of the Hamiltonian with respect to a given quantum state should be a real number. In contrast, for an open physical system, the Hamiltonian operator describing the system may not possess the property of Hermiticity~\cite{moiseyev2011non}, i.e.,
	$\hat{H}\neq \hat{H}^{\dag}$.
	In this case, $\hat{H}$ and $\hat{H}^{\dag}$ have distinct sets of eigenstates, known as right eigenstates and left eigenstates, respectively, 
	\begin{eqnarray}
		\hat{H}\left | \psi^r_n  \right \rangle=E_{n} \left | \psi^r_n  \right \rangle, \quad
		\hat{H}^\dag \left | \psi^l_n  \right \rangle=E^{*}_{n} \left | \psi^l_n  \right \rangle, 
		\label{eq:eigen}
	\end{eqnarray}
	where $E_n^{*}$ is the complex conjugate of $E_n$ and $n$ is the spectral label. In the context of non-Hermitian physics, $\{\psi^r_n\}$and$\{\psi^l_n\}$ are referred to as biorthogonal basis vectors, because a biorthogonal relationship exists between left and right eigenstates~\cite{banach1987theory},
	\begin{eqnarray}
		\left \langle \psi^{l}_{m}  |   \psi^{r}_{n} \right \rangle =c_{n}\delta_{mn},
		\label{eq:biorthogonal}
	\end{eqnarray}
	in which $c_n=\left \langle \psi^{l}_{n}  |   \psi^{r}_{n} \right \rangle $ and $\delta_{mn}$ denotes the Kronecker delta function.
	When the system is in the quantum state $\left | \psi^{r}_{n} \right \rangle$, the expected value of the operator $\hat{A}$ can be expressed as~\cite{moiseyev2011non}
	\begin{eqnarray}
		\langle \hat{A} \rangle=\frac{ \langle \psi_n^l  |\hat{A} | \psi_n^r  \rangle }{\left \langle \psi^{l}_{n}  |   \psi^{r}_{n} \right \rangle} ,
		\label{eq:expectaion_A}
	\end{eqnarray}
	It is worth noting that in the non-Hermitian case, the expectation value of the Hamiltonian cannot be guaranteed to be a real number.
	Computing the energy levels of many-body systems using classical computers becomes increasingly difficult as the system size grows. This is due to the NP-hard problem for solving eigenvalues of generic Hamiltonian, which scales exponentially with the system size~\cite{barahona1982computational}.
	
	Quantum computing has great potential for solving the eigenvalue problem of massive Hamiltonians using the variational quantum eigensolver (VQE) algorithm. VQE is a highly promising quantum algorithm for near-term quantum computers, which can be utilized to compute the ground state and low-lying excited states of a given system. The VQE algorithm employs a parameterized quantum circuit to prepare a trial state, and the expectation value of the Hamiltonian is measured. Subsequently, the parameters of the trial state are optimized iteratively on a classical computer to minimize the cost function, which is designed based on the variational principle in the variational quantum algorithm~\cite{cerezo2021variational}. In VQE, the energy of a system is commonly used to construct the cost function. However, the classical optimizer requires the expectation value to be a real number when minimizing the cost function. Hence, while the VQE algorithm has demonstrated great potential in solving the eigenvalue problem of Hermitian Hamiltonians, it has limitations in solving non-Hermitian physical systems due to the non-real expected value of such systems.
	
	To surmount this obstacle, we introduce a cost function in the variational quantum algorithm, which represents the energy variance of the system. The energy variance is zero if and only if the system is in an eigenstate. Thus, we can get the eigenstates of the system and the corresponding eigenvalues. Our algorithm facilitates the application of the VQE algorithm to the computation of non-Hermitian eigenenergies, enabling its use in a broader range of physical systems. In the following, we provide a detailed exposition of the  algorithm.
	
	\subsection{Variational quantum eigensolver}
	
	\begin{figure}[tp]
		\centering
		\includegraphics[width=0.9\linewidth]{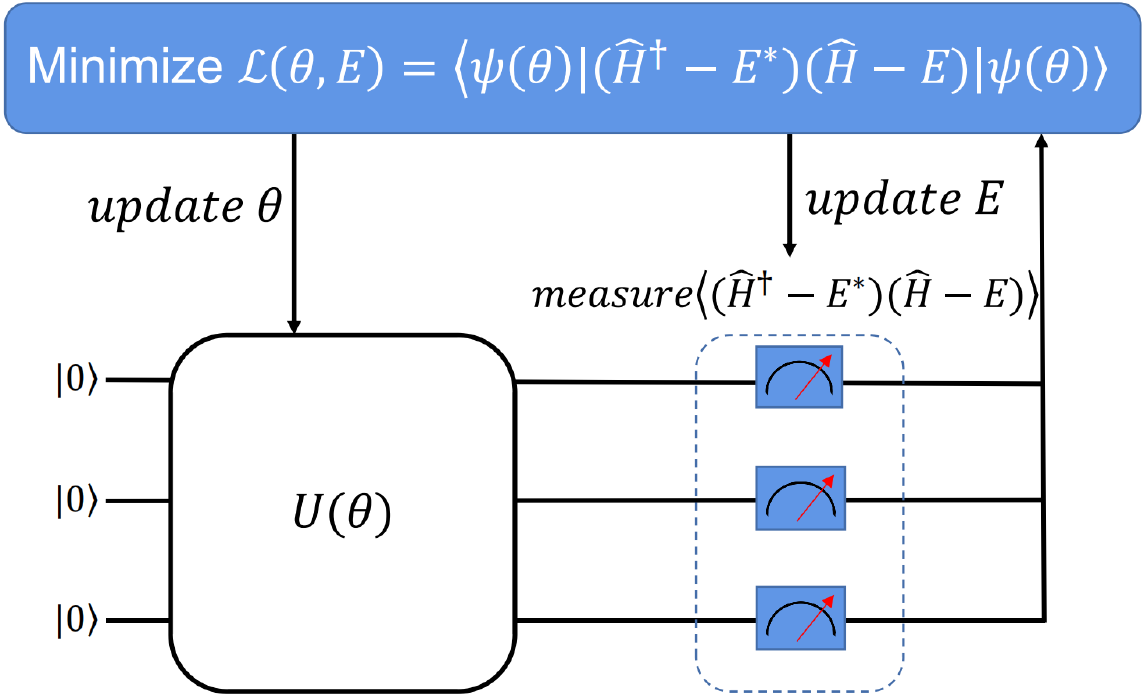}
		\caption{ Illustration of the variational quantum algorithm, which can prepare the eigenstates and compute the corresponding eigenenergies. The quantum circuit $U(\theta)$ is parameterized by $\theta$ and should be performed on a quantum computer. The variable parameters $(\theta,E)$ are updated and optimized with classical computing in order to minimize the cost function. }\label{Fig:illustration}
	\end{figure}
	
	Given an open physical system that can be described by a non-Hermitian Hamiltonian  $\hat{H}$, our aim is to employ quantum variational algorithms to compute the system's eigenvalues and eigenstates. To begin with, we employ the Hermitianization technique to construct a Hermitian Hamiltonian from the given non-Hermitian Hamiltonian $\hat{H}$
	\begin{eqnarray}
		\hat{M}(E)=(\hat{H}^{\dag}-E^{*})(\hat{H}-E),
		\label{eq:M}
	\end{eqnarray}
	where $E^{*}$ is the complex conjugate of $E$. The variable $E$ in the Hamiltonian matrix $M(E)$ is a complex number, which reflects the fact that the eigenvalues of the non-Hermitian Hamiltonian are generally complex. Through the construction method of Eq.~\eqref{eq:M}, we obtain a Hermitian Hamiltonian matrix $M(E)$ that is non-negative. The Hermitian Hamiltonian matrix $M(E)$ is semi-positive definite which satisfies
	\begin{eqnarray}
		\langle \hat{M}(E) \rangle =\left \langle \psi \right |(\hat{H}^{\dag}-E^{*})(\hat{H}-E)\left |\psi \right \rangle \geq 0.
		\label{eq:M_ex}
	\end{eqnarray}
	As the Eq.~\ref{eq:M_ex} shows, the expected value of $\hat{M}$ is actually closely related to the variance energy of the Hamiltonian $\hat{H}$.  The difference is that in $ \langle \hat{M}(E) \rangle$ the energy is parameterized. The condition for $ \langle \hat{M}(E) \rangle$ to be equal to zero is if and only if 
	\begin{eqnarray}
		(\hat{H}-E)\left |\psi \right \rangle = 0.
	\end{eqnarray}
	This condition implies that the state vector $\left |\psi \right \rangle$ is a right eigenstate of the Hamiltonian $\hat{H}$, with the corresponding eigenvalue $E$. Likewise, if we want to solve the left eigenstate, we only need to replace $\hat{M}(E)$ with $\hat{M^{'}}(E)$,
	\begin{eqnarray}
		\hat{M^{'}}(E)=(\hat{H}-E)(\hat{H}^{\dag}-E^{*}).
		\label{eq:M1}
	\end{eqnarray}
	
	The cost function for the variational quantum algorithm can be designed as following, 
	\begin{eqnarray}
		{\cal L} (\theta,E)&&=\left \langle \psi(\theta) \right |(\hat{H}^{\dag}-E^{*})(\hat{H}-E)\left |\psi(\theta) \right \rangle \notag \\
		&&=\left \langle \psi(\theta) \right |\hat{M}(E)\left |\psi(\theta) \right \rangle,
		\label{eq:loss}
	\end{eqnarray}
	where $\left |\psi(\theta) \right \rangle=U(\theta)\left |0 \right \rangle$ and $U(\theta)$ is a unitary operation, which can be implemented with parameterized quantum circuits. As presented in Eq.~\eqref{eq:loss}, the cost function ${\cal L} (\theta,E)$ is equivalent to the expected value of $\hat{M}(E)$ in a specific quantum state $\left |\psi(\theta) \right \rangle$. By decomposing $\hat{M}(E)$ as a sum of Pauli operators and performing each Pauli measurement alone, the cost function can be obtained efficiently. 
	
	The cost function can be minimized with a hybrid quantum-classical optimization method. 
	As depicted in the Fig.~\ref{Fig:illustration}, the $U(\theta)$ is utilized to prepare the quantum state $\left |\psi(\theta) \right \rangle$ on the quantum computer. Subsequently, the expected value of $\hat{M}(E)$ is measured, and classical optimization techniques are applied to minimize the expected value by updating the parameters $(\theta, E)$. Upon achieving a minimum value of the cost function, a right eigenstate $\left |\psi(\theta) \right \rangle$ of $\hat{H}$ and the corresponding eigenvalue $E$ are obtained. 
	
	\subsection{Optimization strategy}
	For a given Hamiltonian $\hat{H}$, there may exist multiple sets of eigenstates and corresponding eigenvalues which makes the cost function reach the minimum value of zero. In general, when minimizing a cost function using a classical optimizer, different initial parameter values can lead to different solutions for the optimization problem.  Therefore, in order to obtain a specific eigenstate, such as the ground state, or the eigenstate with the largest loss, we need to design reasonable optimization schemes. To this end, we adopt a two-step optimization strategy, which consists of two different optimization processes. For convenience, we denote $E$ as $E_r+\textrm{i}E_i$, where $E_r$ is the real component and $E_i$ is the imaginary component. Thus, the cost function ${\cal L}(\theta, E)$ can be represented as ${\cal L}(\theta,E_r,E_i)$.
	
	As described in \algorithmcfname~\ref{alg:algorithm_1}, in the first step, we leave the initial value of $E_r$ unchanged and update the parameters $E_i$ and $\theta$ to minimize the cost function; in the second step, all parameters are updated together to minimize the cost function to 0. Adopting the two-step optimization strategy, we can obtain the eigenstate $\left |\psi(\theta) \right \rangle$ of the Hamiltonian $\hat{H}$ with high accuracy. The real component  of the corresponding eigenvalue $E$ is very close to the initial value $E_{r0}$. To obtain the ground state of the Hamiltonian $\hat{H}$, it is necessary to set the real component of  $E_{r0}$ to a sufficiently small value. This is because the ground state of the Hamiltonian has the  lowest eigenenergy~(concerning the real part of the energy).
	
	To properly set the initial value of $E_r$,  we need to estimate the range of energy levels of the system. Given a general Hamiltonian, it is always possible to express it as the sum of Hermitian and anti-Hermitian components,
		\begin{eqnarray}
			\hat{H}=\hat{H}_r+\textrm{i} \hat{H}_i ,
		\end{eqnarray}
		in which $\hat{H}_r=\sum_j b_j \hat{O}_j$ and $\hat{H}_i=\sum_k d_k \hat{O}_k$. $\{\hat{O}_j\}$ are the tensor product of Pauli operators, and the expected value within the normalized quantum state adheres to the inequality $-1 \leq \langle \hat{O}_j \rangle \leq 1$. Therefore, we can directly obtain the value range of the system energy,
		\begin{eqnarray}\label{Eq:range_E}
			E_{min}^r \leq E_r \leq E_{max}^r ,\quad  E_{min}^i \leq E_i \leq E_{max}^i
		\end{eqnarray}
		where $E_{min}^r=-\sum_j |b_j|, E_{max}^r=\sum_j |b_j|, E_{min}^i=-\sum_k |d_k|$ and $E_{max}^i=\sum_k |d_k|$. According to Eq.\eqref{Eq:range_E}, we can reasonably set the initial values of energy parameters $E_{r0}=E_{min}^r$, which is sufficiently small for solving the ground state with the two-step optimization.
	
	\begin{algorithm}[tb]
		\SetAlgoLined
		\KwIn{  Input the initial energy value, $E_r=E_{r0}, E_i$, and the initial parameter set, $\theta$. }  
		\While{${\cal L}(\theta,E_r,E_i)$ has not converged} {
			$\theta\gets\theta - \alpha \frac{\partial {\cal L}}{\partial \theta} $\;
			$E_i\gets E_i - \alpha \frac{\partial {\cal L}}{\partial E_i} $\;
		}
		\While{${\cal L}(\theta,E_r,E_i)$ has not converged} {
			$\theta\gets\theta - \alpha \frac{\partial {\cal L}}{\partial \theta} $\;
			$E_i\gets E_i - \alpha \frac{\partial {\cal L}}{\partial E_i} $\;
			$E_r\gets E_r - \alpha \frac{\partial {\cal L}}{\partial E_r} $\;
		}
		return $E_r, E_i, \theta$
		\caption{two-step optimization}
		\label{alg:algorithm_1}
	\end{algorithm}
	By utilizing \algorithmautorefname~\ref{alg:algorithm_1}, we can effectively identify an eigenstate whose eigenenergy is in close proximity to the initial value $E_{r0}$. Consequently, by manipulating the initial value, $E_{r0}$, we can systematically traverse the eigenstates of the Hamiltonian. 
	In order to calculate the ground state and the low-lying excited state, we adopted the strategy described in the \algorithmautorefname~\ref{alg:algorithm_3} (see in the Appendix~\ref{App:spectrum})which is based on \algorithmautorefname~\ref{alg:algorithm_1}. It first solves the ground state energy, and then starts from the ground state energy to find the low-lying excited state step by step through gradually increasing the value of $E_{r0}$. This approach enables us to explore the complete spectrum.
	
	Similar to the process for determining the ground state, the two-step optimization approach can also be utilized to determine the eigenstate with the greatest absolute value of the imaginary component of energy. First, we keep the initial value of $E_i$ constant and update the parameters $E_r$ and $\theta$; then, all parameters are updated together to minimize the cost function to 0. Then we can obtain the eigenstate $\left |\psi(\theta) \right \rangle$, and the imaginary component of the corresponding eigenvalue $E$ is in close proximity to its initial value $E_{i0}$. So  we just need to set $E_{i0}$  to a large value, $E_{i0}=E_{max}^i$.

	\subsection{\label{subsec:hadamard}Biorthogonal relations and operator expected values}
	As shown in Eq.\eqref{eq:biorthogonal}, there is a biorthogonal relationship between the left and right eigenvectors of the non-Hermitian Hamiltonian. Under the biorthogonal basis, the expected value of the operator is not directly measurable as in the Hermitian case. In this section, we will describe in detail how to use the Hadamard test to verify the biorthogonal relationship and determined the expected value of the given operator in a eigenstate.
	
	Initially, we employ a variational quantum algorithm to generate both the left eigenstates $\left | \psi^l_n  \right \rangle $ and right eigenstates $\left | \psi^r_n  \right \rangle $ eigenstates of the Hamiltonian $\hat{H}$ on a quantum computing device
	\begin{eqnarray}
		\left | \psi^r_n  \right \rangle =U(\theta^r_n)\left | 0  \right \rangle, \quad
		\left | \psi^l_m  \right \rangle =U(\theta^l_m)\left | 0  \right \rangle,
	\end{eqnarray}
	where $U(\theta_n^l)$ and $U(\theta_n^r)$ are the parameterized quantum circuit associated with the eigenstate $| \psi_n^l \rangle$ and $| \psi_n^r \rangle$, respectively.
	The biorthogonality of the eigenstates of the non-Hermitian Hamiltonian can be verified through the evaluation of fidelity. The fidelity between $\left | \psi^r_n  \right \rangle $ and $\left | \psi^l_n  \right \rangle $ can be described as
	\begin{eqnarray}
		{\cal F}&&=\sqrt{||\left \langle \psi^l_m| \psi^r_n \right \rangle||}=\sqrt{|c_n|^2}\delta_{m,n} \notag\\
		&&=\sqrt{||\left \langle 0|U^\dagger(\theta^l_m)U(\theta^r_n)|0 \right \rangle||}.
		\label{eq:fidelity}
	\end{eqnarray}

	\begin{figure}[tp]
		\centering
		\includegraphics[width=\linewidth]{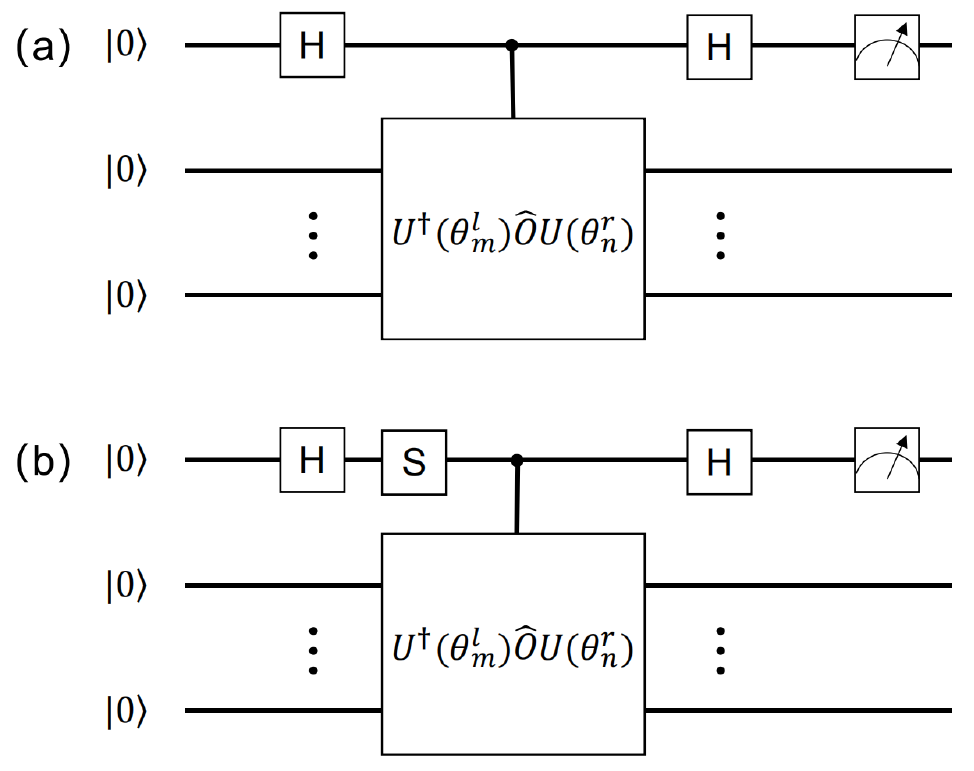}
		\caption{(a) The quantum circuit diagram for the real part of $ \langle 0|U^\dagger(\theta^l_m) \hat{O}U(\theta^r_n)|0 \rangle$ by the Hadamard test, in which H denotes the Hadamard gate; (b) The quantum circuit diagram for  the imaginary part of $ \langle 0|U^\dagger(\theta^l_m) \hat{O} U(\theta^r_n)|0  \rangle$  by the Hadamard test, in which S denotes the $\frac{\pi}{2}$ phase gate. \label{Fig:hadamard_test}}
	\end{figure}

	Suppose that the operator $\hat{A}$ can be expressed as a linear combination of Pauli operators
	\begin{eqnarray}
		\hat{A}=\sum_i a_i \hat{O}_i,
	\end{eqnarray}
	where $\hat{O}_i$ denotes a Pauli operator. 
	According to the Eq.~\ref{eq:expectaion_A}, the expected value of the operator $\hat{A}$ in the eigenstate $| \psi_n^r \rangle$ is given by
	\begin{eqnarray}
		\langle \hat{A} \rangle=\frac{ \langle \psi_n^l  |\hat{A} | \psi_n^r  \rangle }{\left \langle \psi^{l}_{n}  |   \psi^{r}_{n} \right \rangle} 
		=\frac{ \sum_i a_i \langle \psi_n^l  |\hat{O}_i | \psi_n^r  \rangle }{\left \langle \psi^{l}_{n}  |   \psi^{r}_{n} \right \rangle} .
		\label{eq:exp_A}
	\end{eqnarray} 
	
	As shown in Eq.~\ref{eq:fidelity} and Eq.~\ref{eq:exp_A}, to figure out the fidelity and the expected value of the operator $\hat{A}$, we just have to obtain the value of $\left \langle \psi^l_m| \psi^r_n \right \rangle$ and $  \langle \psi_n^l  |\hat{O}_i | \psi_n^r  \rangle$.
	As portrayed in Fig.~\ref{Fig:hadamard_test}(a), the quantum circuit at hand possesses the capability to obtain the real component of $  \langle \psi_n^l  |\hat{O} | \psi_n^r  \rangle$ through the Hadamard test. Upon performing a measurement on the first qubit, the probability of finding the system in the state $\left | 0 \right \rangle$ is given by~\cite{cleve1998quantum}
	\begin{eqnarray}
		P_r(0)&&=\frac{1}{2}+\frac{1}{2} \Re ( \langle 0|U^\dagger(\theta^l_m) \hat{O} U(\theta^r_n)|0  \rangle ) \notag \\
		&&=\frac{1}{2}+\frac{1}{2} \Re ( \langle \psi_m^l  |\hat{O} | \psi_n^r  \rangle ).
	\end{eqnarray}
	
	On the other hand, as depicted in Figure~\ref{Fig:hadamard_test} (b), the aforementioned quantum circuit can employ the Hadamard test to obtain the imaginary part of $\left \langle 0|U^\dagger(\theta^l_m)U(\theta^r_n)|0 \right \rangle$. The probability of observing the state $\left | 0 \right \rangle$ after measuring the first qubit can be expressed as follows~\cite{PhysRevA.104.042418}
	\begin{eqnarray}
		P_i(0)&&=\frac{1}{2}-\frac{1}{2}\Im ( \langle 0|U^\dagger(\theta^l_m)\hat{O}U(\theta^r_n)|0  \rangle ) \notag \\
		&&=\frac{1}{2}-\frac{1}{2}\Im( \langle \psi_m^l  |\hat{O} | \psi_n^r  \rangle ).
	\end{eqnarray}
	Thus, the value of $ \langle \psi_m^l  |\hat{O} | \psi_n^r  \rangle $ can be obtained ,
	\begin{eqnarray}
		\langle \psi_m^l  |\hat{O} | \psi_n^r  \rangle &&=\Re( \langle \psi_m^l  |\hat{O} | \psi_n^r  \rangle ) +\textrm{i}~\Im ( \langle \psi_m^l  |\hat{O} | \psi_n^r  \rangle ) \notag \\&&=2P_r(0)-1+\textrm{i}~[~2-P_i(0)].
	\end{eqnarray}
	The overlapping $\langle \psi_m^l  | \psi_n^r \rangle$ can be obtained by setting $\hat{O}=\hat{I}$. 
	
	\begin{figure}[tp]
		\centering
		\includegraphics[width=\linewidth]{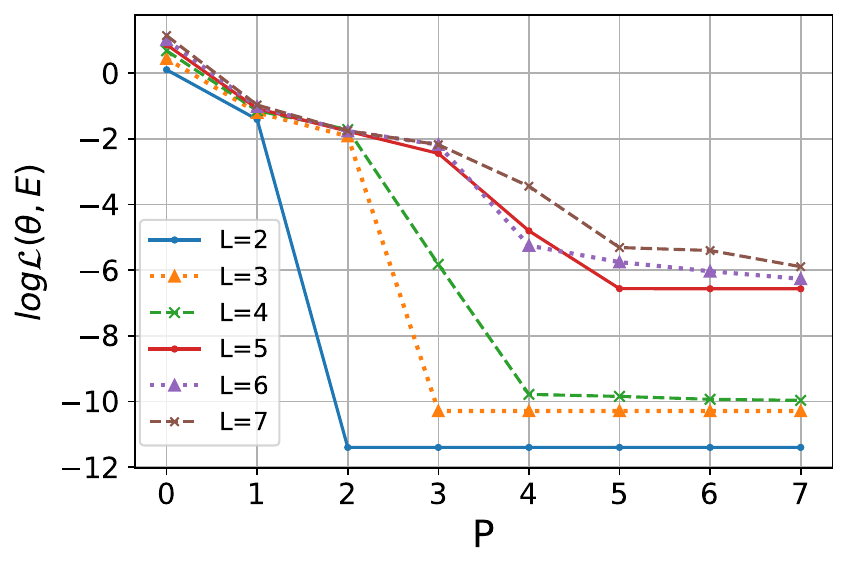}
		\caption{ The logarithm of the cost function ${\cal L}(E,\theta)$ as a function
			of the circuit depth $P$, for various system sizes $L$. Here, $\lambda=1, \kappa=0.8$.  }\label{Fig:loss_P}
	\end{figure}
	\section{\label{sec:sumulaton} SIMULATION RESULTS}
	In this section, we will utilize a standard non-Hermitian Hamiltonian as a demonstration of our algorithm.  To investigate the performance of our algorithm, we conduct numerical simulations under different conditions. Additionally, we assess the algorithm's practicality on Noisy Intermediate-Scale Quantum  device by considering the effects of quantum noise and utilizing error mitigation techniques to enhance its accuracy. We carried out the numerical simulations using the open-source package qibo\cite{efthymiou2021qibo} and qutip~\cite{johansson2012qutip}.

	\begin{figure}[tp]
		\centering
		\includegraphics[width=\linewidth]{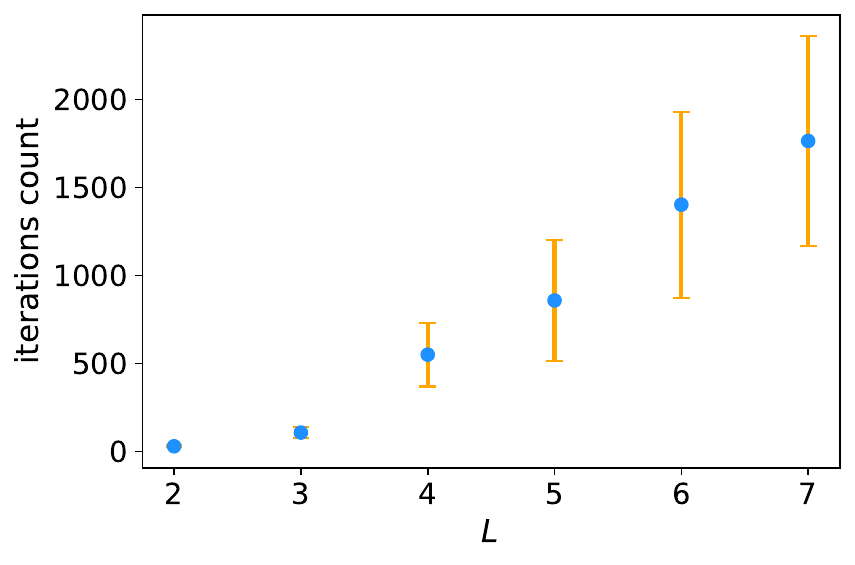}
		\caption{ The variability of iterations in proportion to system size.
			As the system size expands, there is a corresponding increase in the requisite number of iterations for the loss function to converge to its minimum value. For various system sizes $L$, we set the depth of the quantum circuit $P$ equal to $L$ in order to ensure the convergence of the loss function to zero..Here, $\lambda=1, \kappa=0.8$.  }\label{Fig:itercount_L}
	\end{figure}
	\subsection{Performance of quantum algorithm}

	The selected non-Hermitian lattice model is the Ising quantum spin chain in the presence of a magnetic field in the $z$-direction as well as a longitudinal imaginary field~\cite{von1991critical}. This model can be described by the following Hamiltonian
	\begin{eqnarray}
		H_{\lambda,\kappa}=-\frac{1}{2}\sum_j^L (\lambda \sigma_j^x\sigma_{j+1}^x+\sigma_j^z+i \kappa \sigma_j^x),
	\end{eqnarray}
	where $\lambda,\kappa \in R $.
	To implement the Hamiltonian of interest, we device a unitary quantum circuit, which is composed of a series of single qubit rotation gates and two-qubit rotation gates. The design of the circuit is as follows ~\cite{sousa2006universal,PhysRevD.106.054509}
	\begin{eqnarray}
		U(\theta)&&=\prod_{j=1}^{P}U_j(\theta_j) ,\notag\\
		U_j(\theta_j)&&=e^{-i H_{xx}(\alpha_j)}e^{-i H_{z}(\beta_j)}e^{-i H_{x}(\gamma_j )},
	\end{eqnarray}
	where $H_{xx}(\alpha_j)=\sum_l \alpha_{j,l}X_lX_{l+1}$, $H_{z}(\beta_j)=\sum_l \beta_{j,l}Z_l$, $H_{x}(\gamma_j)=\sum_l \gamma_{j,l}X_l$ and $\theta_j = (\alpha_j, \beta_j, \gamma_j)$ is the parameter set used to control the rotation angle of quantum gates.
	
	\begin{figure}[tp]
		\centering
		\includegraphics[width=0.85\linewidth]{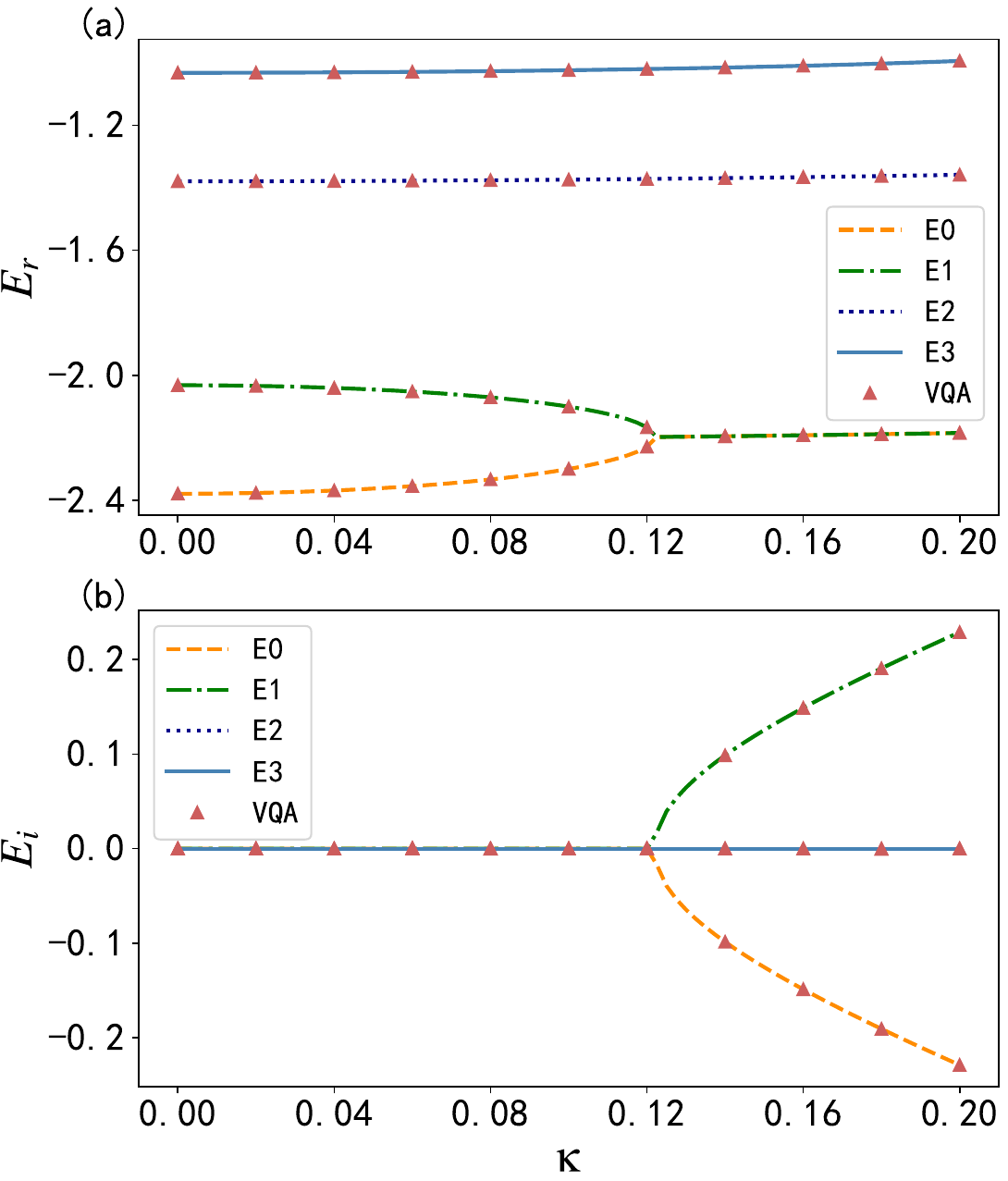}
		\caption{The energy levels of $H_{\lambda,\kappa}$ as a function of $\kappa$, in the case $L=4,\lambda=1$. (a) illustrates the real component of the energy levels, while (b) depicts the imaginary part of the energy levels. Those lines denoted by $E_0$, $E_1$, $E_2$ and $E_3$ are obtained by exact diagonalization. }\label{Fig:spetrum}
	\end{figure}
	
	Fig.~\ref{Fig:loss_P} demonstrates that the accuracy of the computed cost function by our quantum algorithm improves as the depth of the quantum circuit $P$ increases, ultimately approaching zero. This indicates that the obtained eigenstates and corresponding eigenvalues become increasingly accurate. The observed improvement can be attributed to the increased complexity and expressiveness of the circuit, which facilitate a more faithful encoding of the target Hamiltonian's properties~\cite{PRXQuantum.2.040309}.
	
	\begin{table*}[ht]
		\renewcommand{\arraystretch}{1.2}
		\caption{\label{tab:fidelity} Fidelity between the right eigenstates $\left | \psi^{r}_{n} \right \rangle$ and the left eigenstates $\left | \psi^{l}_{n} \right \rangle$, with $\kappa=0.4$ and $\kappa=0.2$,respectively. Here $L=3,P=3,\lambda=1$.
		}
		\begin{ruledtabular}
			\begin{tabular}{ccccccccc}
				&$n=0$ &$n=1$ &$n=2$ &$n=3$ &$n=4$ &$n=5$ &$n=6$ &$n=7$\\
				\hline
				VQA($\kappa=0.4$) &0.7988	&0.7988	&0.9165	&0.7719	&0.7719	&0.9165	&0.7978	&0.7994\\
				exact &0.7988	&0.7988	&0.9165	&0.7719	&0.7719	&0.9165	&0.7979	&0.7994\\
				\hline
				VQA($\kappa=0.2$) &0.1680	&0.1680	&0.9798	&0.5822	&0.5821	&0.9798	&0.9537 &0.9541\\
				exact &0.1681	&0.1681	&0.9798	&0.5821	&0.5821	&0.9798	&0.9537	&0.9541\\
				
			\end{tabular}
		\end{ruledtabular}
	\end{table*}
	The number of iterations is an important metric for variational algorithms. To investigate the relationship between the iteration count and the system size, we employ the quantum algorithm to compute the ground state for systems of various sizes. The closer the initial state $|\psi_0\rangle=U(\theta_0)|0\rangle$ is to the target state, the quicker the quantum algorithm can achieve the desired outcome, resulting in fewer required iterations. Consequently, different initial parameters lead to varying iteration counts. To ensure the reliability, we conduct $20$ numerical simulations and take an average of their outcomes, where the initial parameters $\theta_0$ for each simulation are randomly generated. As illustrated in Fig.~\ref{Fig:itercount_L}, the required number of iterations for the loss function converging to its minimum value increases with an increasing of the system size.

	We utilize the quantum variational algorithm to evaluate the energy spectrum of $H_{\lambda,\kappa}$, as shown in Fig.~\ref{Fig:spetrum}. To compute the ground state and low-lying excited states, we employ the method outlined in the \algorithmautorefname~\ref{alg:algorithm_3}. This involves initially obtaining the ground state energy and subsequently using it as a starting point to determine the low-lying excited states step by step. The outcomes exhibit strong conformity with the corresponding exact values, thereby providing compelling evidence for the effectiveness of our algorithm in computing non-Hermitian Hamiltonian eigenvalues. Additionally, the data presented in Fig.~\ref{Fig:spetrum}(a) manifests a gradual convergence of the energy levels for both the ground and the first excited states with increasing values of $\kappa$, which culminates at the exceptional point. Furthermore, Fig.~\ref{Fig:spetrum}(b) discloses that beyond the exceptional point, both the ground and the first excited states feature the presence of imaginary components in their energy levels, indicating the occurrence of a real-to-complex spectral transition at the exceptional point.
	
	\begin{figure}[tp]
		\centering
		\includegraphics[width=\linewidth]{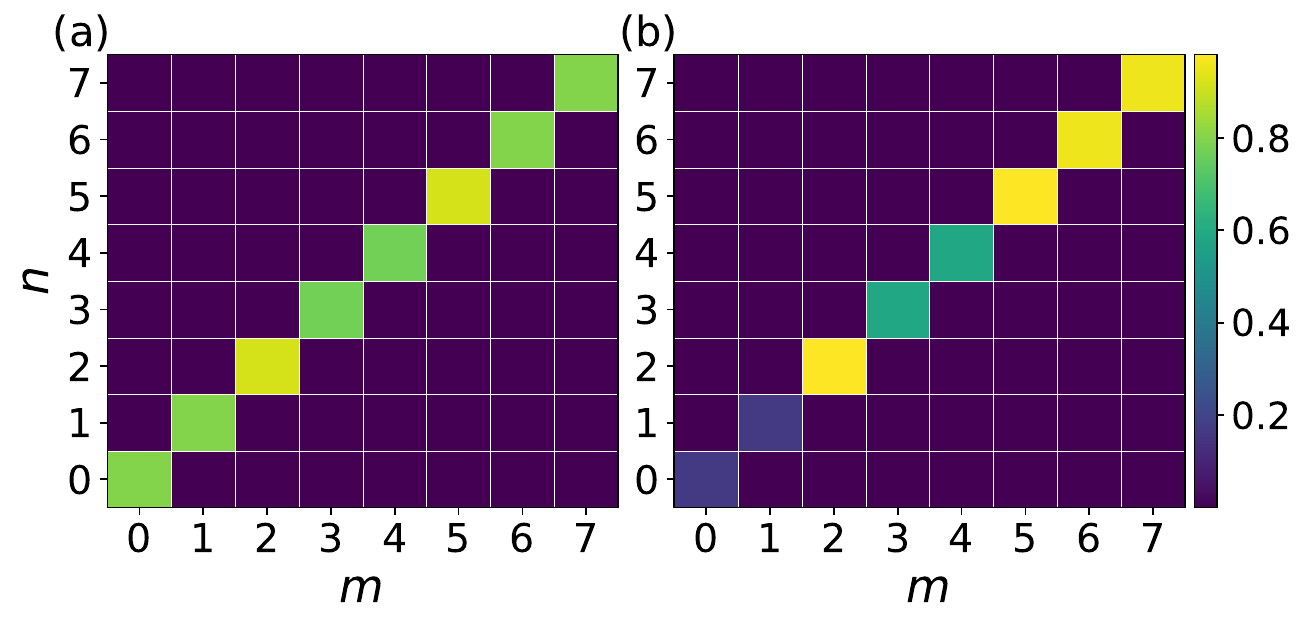}
		\caption{The fidelity between $\left | \psi^r_n  \right \rangle $ and $\left | \psi^l_n  \right \rangle $ for $\kappa=0.4$ in (a) and $\kappa=0.2$ in (b). Here $L=3, \lambda=1$. 
		}\label{Fig:fidelity}
	\end{figure}

	\begin{figure}[bp]
		\centering
		\includegraphics[width=1\linewidth]{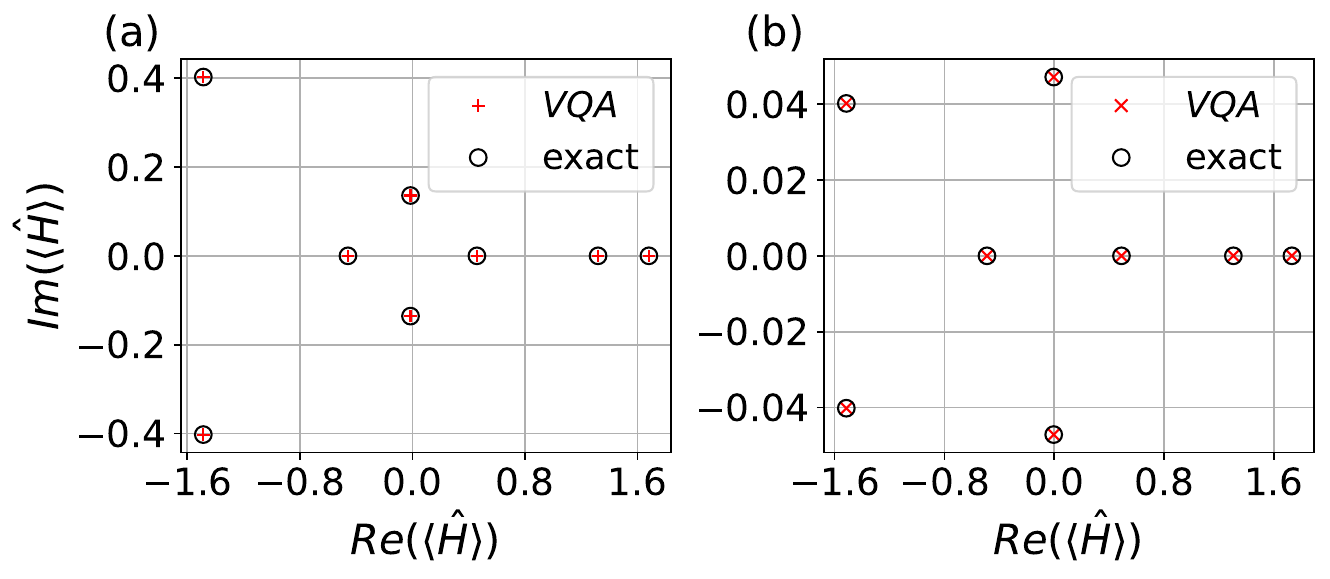}
		\caption{The expected value  of the Hamiltonian operator $H_{\lambda,\kappa}$ according to Eq.~\eqref{eq:exp_A} for $\kappa=0.4$ in (a) and $\kappa=0.2$ in (b). Here $L=3, \lambda=1$.
		}\label{Fig:H_ex}
	\end{figure}

	Using our variational quantum algorithm, we are able to prepare both the right and left eigenstates of a given Hamiltonian on a quantum device. Utilizing the Hadamard test, we can calculate the fidelity between the two eigenstates. As shown in Fig.~\ref{Fig:fidelity}, we have obtained the fidelity between the left and right eigenstates of the Hamiltonian $H_{\lambda,\kappa} $ with $L=3$, which confirms the biorthogonal relationship between the eigenstates of the non-Hermitian Hamiltonian. As can be seen from table~\ref{tab:fidelity}, the results obtained by the quantum algorithm are in good agreement with the exact values. This verifies that our quantum algorithm is suitable for studying non-Hermitian physical systems.
	
	In Section~\ref{subsec:hadamard}, we introduce an approach to determine the expected value of the operator by means of the Hadamard test. To ascertain the soundness of our methodology, we proceed to evaluate the expected value of the Hamiltonian operator $\hat{H} =H_{\lambda,\kappa}$ under biorthogonal base vectors. As depicted in the Fig~\ref{Fig:H_ex}, our quantum algorithm yields results that are in excellent agreement with the exact values, thus attesting to the efficacy of our algorithm.
	
	\begin{figure}[tp]
		\centering
		\includegraphics[width=1\linewidth]{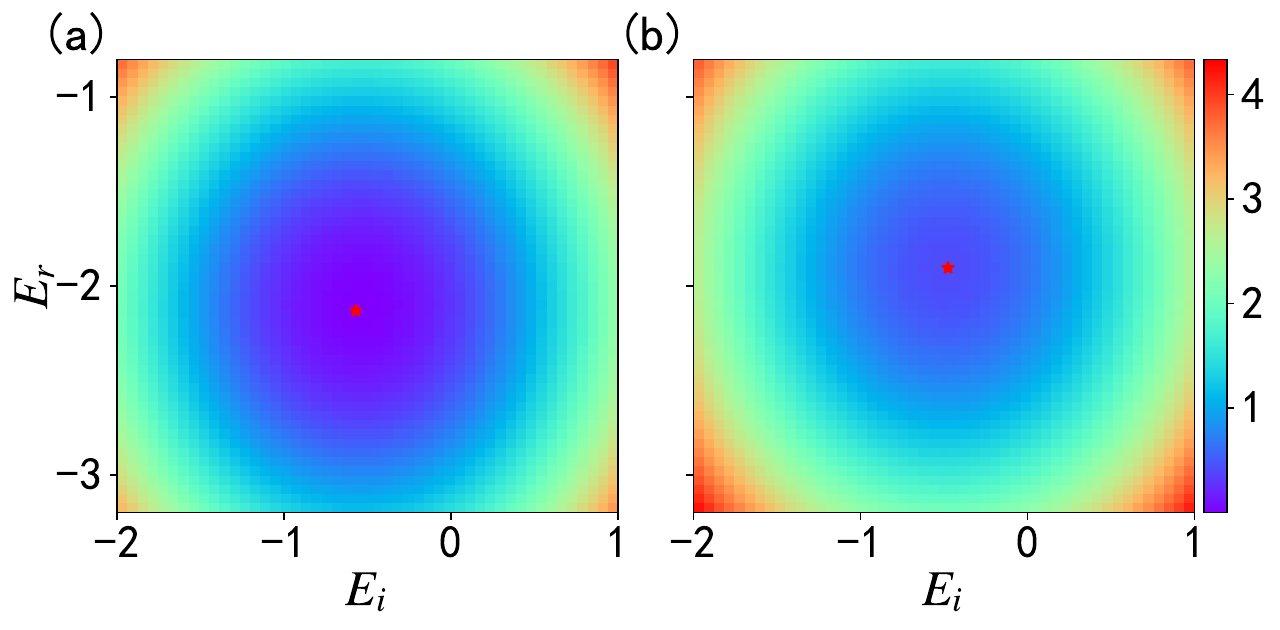}
		\caption{Illustration of effects of quantum noises on the landscapes for the real and imaginary part of the energy parameter. The left and right plots depict the landscape without noise and under the depolarization noise, respectively. The red star in the figure is the minimum value of the loss function, which corresponds to the optimal energy parameter. A comparison shows that noise will lead to a shift of the optimized energy. }
		\label{Fig:landscape}
	\end{figure}

	\subsection{Error mitigation}
	Quantum noise is a major challenge for implementing quantum algorithms on current quantum processors~\cite{Preskill2018quantumcomputingin}. To evaluate and optimize the quamtun algorithms, it is crucial to consider the noise impact in numerical simulations. For demonstration, we adopt a noise model with depolarization(analysis of other noise models as well as the error mitigation can be seen in the Appendix~\ref{App:noise}). The quantum circuit consists primarily of single-qubit quantum gates and two-qubit quantum gates. 
	Therefore, in the numerical simulations, after applying a single-qubit gate, noise in the form of depolarization can be added to each qubit with a probability of $p_1$. 
	\begin{eqnarray}
		\varepsilon(\rho)=(1-p_1 )\rho+\frac{p_1}{3}(X\rho X+Y\rho Y+ Z\rho Z).
		\label{eq:}
	\end{eqnarray}
	Similarly, after applying a two-qubit gate, depolarization noise can be added to each qubit with a probability of $p_2$.
	
	In order to study the influence of noise on quantum algorithms, we set the noise rate of single-qubit quantum gate $p_1=0.001$, and the rate tof noise of two-qubit quantum gate $p_2=0.01$. Since the cost function involves many parameters, it is a great challenge to reveal the individual impact of each parameter. To showcase the influence of noise on the cost function landscape, we specifically chose the energy parameters $E_r$ and $E_i$ as variables. These two parameters not only play a role in the loss function but also represent the system's energy, a crucial quantity we want to investigate. It is useful for investigating how quantum noises can affect the optimized energy. In Fig.~\ref{Fig:landscape}, we illustrate the variations of the cost function with respect to $E_r$ and $E_i$, both in the absence of noise and in the presence of noise. 
	The impact of noise exposure on the cost function landscape is not substantial. Nevertheless, the landscape's overall elevation resulting from such exposure prevents the attainment of a minimum loss value of zero. It is worth noting that the optimal solution for energy may undergo some variations in the presence of noise. This suggests that when implementing our algorithm on a noisy quantum device, the obtained results are likely to deviate to some extent from the ideal ones in the noiseless case. 
	
	\begin{figure}[tp]
		\centering
		\includegraphics[width=\linewidth]{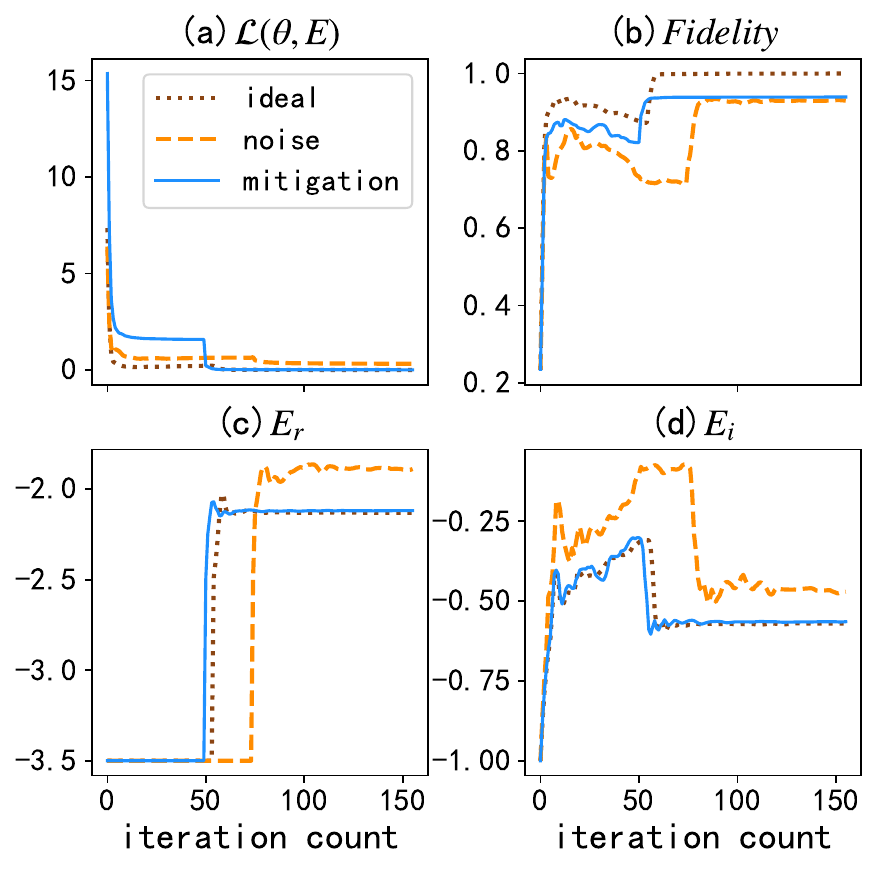}
		\caption{Comparison of optimization processes for ideal, depolarizing noisy and mitigated variational quantum algorithm. The figure illustrates the impact of depolarization noise on quantum algorithms and the effectiveness of noise mitigation techniques. (a) The cost function as a function of the number of iterations; (b) The fidelity respect to target ground state as a function of the number of iterations; (c) The real component of the energy $E$ as a function of the number of iterations; (d) The imaginary component of the energy $E$ as a function of the number of iterations. In all cases $L=4,P=4,\lambda=1,\kappa=0.4$.}
		\label{Fig:iteration}
	\end{figure}
	
	In order to enhance the performance of quantum algorithms on noisy quantum devices, we utilize an error-mitigation technique known as Richardson's deferred method~\cite{richardson1927viii,PhysRevLett.119.180509}, which does not need extra quantum resources and can significantly reduce the error in the expected value of the observation caused by quantum noise. 
	As depicted in Fig. \ref{Fig:iteration}, we investigate the variations of multiple variables in the optimization process with respect to the number of iterations. Specifically, we observe that in the presence of noise, the cost function fails to converge to zero due to the elevation of the cost function landscape caused by noise, as illustrated in Fig. ~\ref{Fig:iteration}(a). To address this issue, we apply error mitigation techniques, which results in a smaller cost function value that approaches the ideal case, indicating an improvement in the algorithm performance. Our findings are further supported by Fig.~\ref{Fig:iteration}(b) and Fig. \ref{Fig:iteration}(c), which demonstrate that the energy eigenvalues obtained using error mitigation techniques are consistent with those in the ideal case. Although error mitigation techniques can reduce the adverse effects of quantum noise on energy measurements, they cannot entirely eliminate quantum noise, and thus there is no significant improvement in the fidelity between the resulting eigenstates and those in the ideal case.
	
	\vspace{0.5cm}
	
	\section{\label{sec:conclusion}conclusion}
	In conclusion, we have proposed a variational quantum algorithm for solving the the eigenvalues and eigenstates of non-Hermitian Hamiltonian by utilizing the zero-variance variational principle. We have also developed a two-step optimization method to efficiently compute specific eigenvectors and eigenvalues. Through numerical simulations, we have demonstrated the effectiveness of our algorithm in computing the eigenvalues of non-Hermitian Hamiltonian and estimating the expected value of operator for non-Hermitian systems. Moreover, we have investigated the impact of quantum noise on our algorithm and incorporates error mitigation techniques to improve its performance. Overall, our work showcases the feasibility for simulating non-Hermitian many-body physics on near-term quantum computers. 
	
	\begin{acknowledgments}
		This work was supported by the National Natural Science Foundation of China (Grant No.12375013 and  No.12275090), the Guangdong Basic and Applied Basic Research Fund (Grant No.2023A1515011460),  and the Guangdong Provincial Key Laboratory (Grant No. 2020B1212060066). 
		
	\end{acknowledgments}


	\appendix
	\section{\label{App:spectrum}Spectrum scanning}
	\begin{algorithm}[bp]
		\SetAlgoLined
		\KwIn{$E_{r}$, $E_{i}$, $\theta$, $n$, $\delta E$, $step$}
		\tcp*[h]{to solve the ground state energy, set $E_{r}$ to a sufficiently small value}\;
		($E_r, E_i, \theta$)=\algorithmautorefname~\ref{alg:algorithm_1}($E_r, E_i, \theta$) \;
		output($E_r, E_i, \theta$) \;
		\tcp*[h]{first, solve the ground state energy}\;
		$i=1$ \;
		\While{$i\leq n$}{
			$E_{temp}=E_r$ \;
			$E_{r}=E_r+\delta E$ \;
			($E_r, E_i, \theta$)=\algorithmautorefname~\ref{alg:algorithm_1}($E_r, E_i, \theta$) \;
			\eIf{$E_r\neq E_{temp}$}{
				output($E_r, E_i, \theta $) \;
				\tcp*[h]{solve the $i-th$ excited state energy}\;
				$i=i+1$ \;
				$\delta E=step$ \;
				\If{$E_i> 0$}{
					$E_i=-E_i$\;
					\While{${\cal L}(\theta,E_r,E_i)$ has not converged} {
						$\theta \gets \theta - \alpha \frac{\partial {\cal L}(\theta,E_r,E_i)}{\partial \theta} $\;
					}
					output($E_r, E_i, \theta $) \;
					\tcp*[h]{solve the $i-th$ excited state energy}\;
					$i=i+1$ \;
					$\delta E=step$ \;
				}
			}{
				$\delta E=\delta E+step$\;
			}
		}
		\caption{spectrum scanning}
		\label{alg:algorithm_3}
	\end{algorithm}
	Using \algorithmautorefname~\ref{alg:algorithm_1}, we can efficiently locate an eigenstate with an eigenenergy that is close to the initial estimate $E_{r0}$. Therefore, by adjusting the initial estimate, $E_{r0}$, we can systematically explore the eigenstates of the Hamiltonian.
	To compute the ground state and the low-lying excited states, we employed the strategy outlined in the \algorithmautorefname~\ref{alg:algorithm_3} based on \algorithmautorefname~\ref{alg:algorithm_1}. It first determines the ground state energy, and then iteratively searches for the low-lying excited states by incrementing the value of $E_{r0}$ from the ground state energy. This approach enables us to obtain the complete spectrum.
	
	\section{\label{App:noise} Noise and the error mitigation }
	\subsection{Bit-flip noise}
	\begin{figure}[bbbp]
		\centering
		\includegraphics[width=\linewidth]{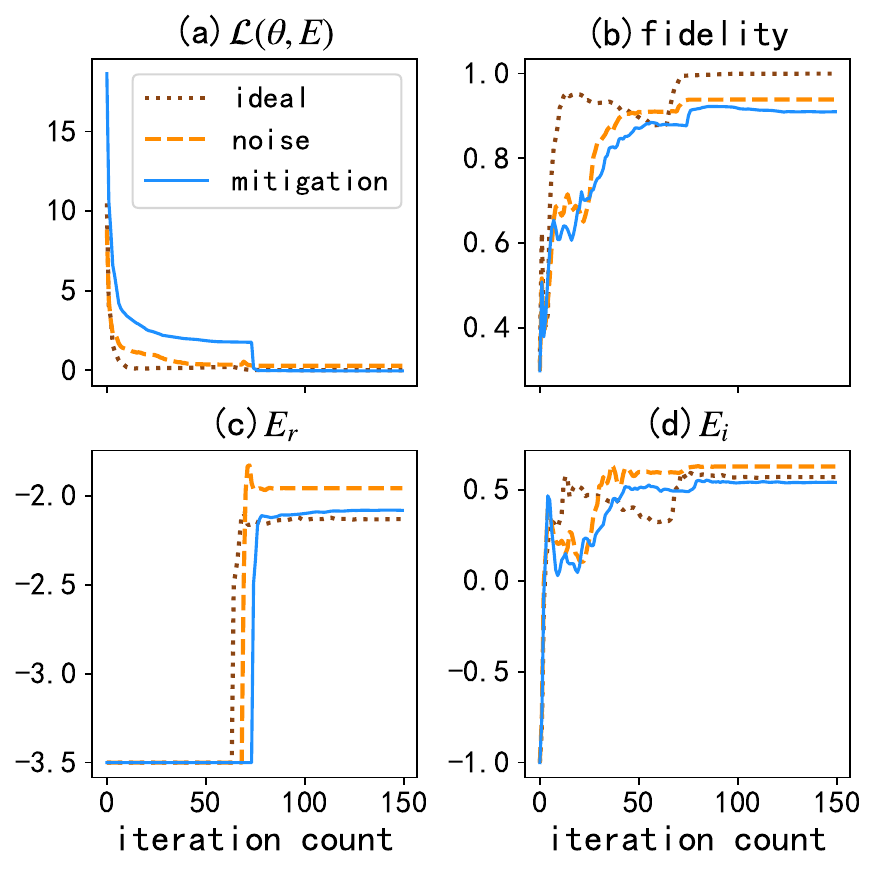}
		\caption{Bit-flip noise and its error mitigation. In (a),(b),(c),(d), the dependence of the cost function, the fidelity, the real and the imaginary components of the energy with the iteration count are shown respectively. In all cases $L=4,P=4,\lambda=1,\kappa=0.4$.}
		\label{Fig:iteration_bt}
	\end{figure}
	
		The bit-flip noise on a qubit can be represented as:
		\begin{equation}
			\varepsilon_1(\rho) = (1-p)\rho + pX\rho X,
		\end{equation}
		where $p$ is the probability of noise occurrence.
		In our numerical simulations, we set the noise probability for a single-qubit gate as $p_1=0.003$, and for a two-qubit gate as $p_2=0.03$, aiming to investigate the impact of noise on quantum algorithms.

	As illustrated in Fig.\ref{Fig:iteration_bt}, the energy eigenvalues generated by the quantum algorithm display evident variations in the presence of bit-flip noise, deviating from the scenario without noise. Additionally, the fidelity of the prepared quantum eigenstates undergoes a reduction. 
	After applying error mitigation techniques, the energy eigenenergies obtained exhibit a heightened level of closeness to the ideal condition. This shows the efficacy of the error mitigation approach in alleviating the negative impact induced by bit-flip noise. Because noise mitigation techniques cannot completely eliminate the noise of quantum devices, the fidelity of the eigenstates obtained by quantum algorithms has not been improved.

	\subsection{Phase-flip noise}
	\begin{figure}[tttp]
		\centering
		\includegraphics[width=\linewidth]{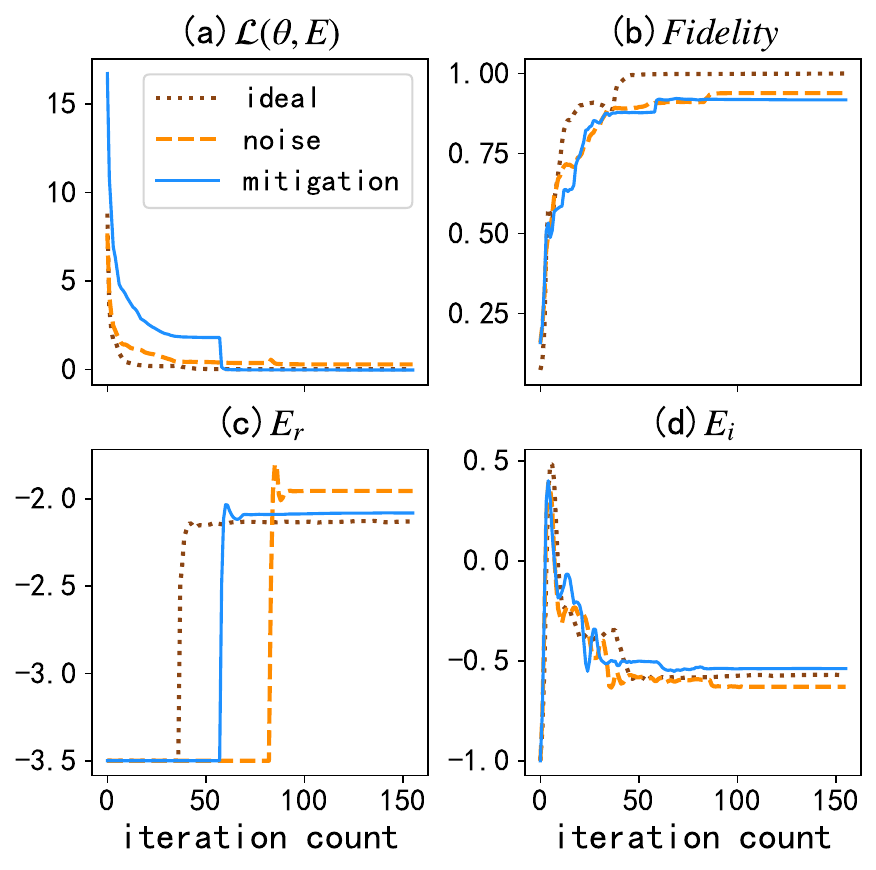}
		\caption{Phase-flip noise and its error mitigation. All other setups are the same as Fig.~\ref{Fig:iteration_bt}}
		\label{Fig:iteration_ph}
	\end{figure}
	
	Another important type of quantum noise is phase-flip noise. The phase-flip transforms the qubit state $|\phi\rangle=\alpha_0 |0\rangle+\alpha_1 |1\rangle$ into $|\psi \rangle=\alpha_0 |0\rangle-\alpha_1 |1\rangle$, with a probability $p$. The noise of phase-flip on a qubit can be written as, 
		\begin{eqnarray}
			\varepsilon_1(\rho)=(1-p )\rho+p Z\rho Z,
		\end{eqnarray}
		In our numerical simulations, we set the noise probability $p_1=0.003$ for the single-qubit gate and $p_2=0.03$ for the two-qubit gate.

	
    As shown in the Fig.\ref{Fig:iteration_ph}, under the influence of phase flip noise, the fidelity between the prepared quantum eigenstate and the target state  has declined. Meanwhile, compared with the situation without noise, the energy eigenvalues obtained by quantum algorithms exhibit a certain degree of deviation under the influence of noise. Error mitigation techniques can reduce measurement errors, but cannot eliminate the noise of quantum devices. Therefore, after using error mitigation techniques, the energy eigenvalues obtained by quantum algorithms are more accurate, as shown in Fig.\ref{Fig:iteration_ph} $(c)$ and $(d)$. However, the fidelity of the prepared eigenstates has not been improved.
	
	In summary, error mitigation techniques can assist us in obtaining more accurate energy eigenvalues, but improving the fidelity of the eigenstates remains challenging. We look forward to the development of more advanced quantum computers in the future, where our algorithm can perform even more effectively.


	\bibliographystyle{apsrev4-2}
	\bibliography{ref}
	
	
\end{document}